\documentclass[aps,prl,reprint]{revtex4-2}
\usepackage{comment}
\usepackage{xcolor}
\usepackage{multirow}
\usepackage{graphics,graphicx,epsfig}
\usepackage[caption=false]{subfig}
\usepackage{amssymb,amsmath}
\usepackage{epstopdf}
\usepackage{amsmath}
\usepackage{bbold}
\usepackage{mathtools}
\usepackage{hyperref}
\hypersetup{colorlinks=true, citecolor = blue, linkcolor=blue, filecolor=magenta, urlcolor=blue}

\begin{document}

\title{Multipartite quantum synchronization in driven-dissipative spin networks}

\author{Jatin Ghildiyal}
\email{jatin.20phz0017@iitrpr.ac.in}
\author{Shubhrangshu Dasgupta}
\author{Asoka Biswas}
\affiliation{Department of Physics, Indian Institute of Technology Ropar, Rupnagar, Punjab 140001, India}

\begin{abstract}

Traditional studies of quantum synchronization in finite-dimensional spin systems usually focus on entrainment of a single spin to an external drive, leaving mutual synchronization among multiple spins poorly understood. We investigate mutual quantum synchronization across bipartite and tripartite spin architectures in a driven-dissipative cavity-coupled quantum Rabi network in the dispersive adiabatic regime. Inspired by the Kuramoto model of classical phase synchronization, we introduce a phase-space measure of quantum synchronization using the first angular moments of Husimi-$Q$ phase distributions. To compare the collective phase localization with reduced-subsystem contributions and local individual phase locking, we construct a bipartite and tripartite synchronization gap. For two spins, the steady-state synchronization gap is negative. This reflects that the relative phase is localized more strongly than the individual spin phases can account for, signaling mutual synchronization beyond independent entrainment by the drive. For three spins in a triangular network, the synchronization gap is positive in the parameter regime studied, i.e., the collective three-body localization is bounded by the pairwise measures. Our results underscore that quantum phase synchronization strongly depends upon network topology and establish the phase-sensitive synchronization gaps as an effective diagnostic for multipartite phase correlations in open quantum systems.

\end{abstract}

\maketitle

Quantum synchronization refers to the emergence of phase-locked collective oscillations in coupled dissipative quantum systems driven out of equilibrium. Unlike classical synchronization, which is characterized by trajectory convergence in phase space, the quantum regime lacks a globally valid phase-space description due to quantum fluctuations. In this regime, one therefore requires operational diagnostics, e.g., two-time correlation functions, synchronization error measures, mutual information, and phase-space quasiprobability distributions~\cite{Giorgi2012, LeeSadeghpour2013, Walter2014}.

Most investigations of quantum synchronization have focused on continuous-variable systems or systems possessing an oscillator-like limit cycle. On the other hand, in finite-dimensional spin systems, synchronization is frequently discussed as the entrainment of an individual spin to an externally applied coherent drive. An open question, however, lies in whether two or more interacting spins can establish a well-defined phase relationship with each other. While the external drives and dissipative channels maintain the system away from equilibrium, mutual synchronization is expected to be identified through phase correlations developed internally among the spin subsystems.

This distinction is particularly important for spin-\(\tfrac12\) systems. A spin does not support a conventional oscillator phase or a classical limit-cycle trajectory \cite{Ye1986Theory,PhysRevLett.121.053601}. Nevertheless, the azimuthal coordinate of its Bloch sphere provides a natural representation of the phase associated with the spin coherence. Spin-coherent states and the corresponding Husimi-\(Q\) quasiprobability distributions can consequently describe the angular statistics of individual and composite spin states \cite{Roulet2018,PhysRevLett.121.053601}. Localization of a single-spin Husimi distribution $Q_j(\theta_j,\phi_j)$ in its azimuthal coordinate $\phi_j$ ($\theta_j$ being the polar coordinate) characterizes phase locking with respect to an external reference. By contrast, localization of a joint distribution $Q_{12}\!\left(\theta,\phi_-^{(12)}\right)$ in a relative-phase coordinate $\phi_-^{(12)}=(\phi_1-\phi_2)/2$ characterizes synchronization between the spins. A joint phase distribution that is uniform in \(\phi_-^{(12)}\) contains no preferred relative phase. Localization around a particular value of \(\phi_-^{(12)}\), on the other hand, signifies spin–spin phase locking. For spins coupled through bosonic fields or common environments, synchronization manifests through the persistence of phase correlations encoded in the off-diagonal elements of the reduced density matrix. The Bloch vector components of individual spins exhibit sustained oscillations whose relative phases encode mutual synchronization, even in the nonequilibrium steady state. We note that in \cite{PhysRevA.105.062206}, two-spin synchronization has been defined in a restricted phase space, which does not essentially extract the phase-locking dynamics among the spins.

To our knowledge, existing phase-space approaches do not
provide a measure of relative-phase localization that extends to three or
more spins, nor a way to compare the collective synchronization with the
contributions of reduced subsystems. Here we address both: we define
relative-phase measures for two and three spins from the first angular
moments of Husimi-$Q$ distributions, and construct synchronization gaps that test
whether the collective synchronization is accounted for by reduced
one- and two-spin states. In this regard, we introduce a class of phase-space synchronization for bipartite and tripartite spin systems, constructed from the ensemble average of the phase $R_j=\left|\langle e^{i\phi_j}\rangle\right|$ and $R_{12}=\left|\left\langle e^{i\phi_-^{(12)}}\right\rangle\right|$.
These quantities measure relative phase localization on the Bloch sphere and provide a natural finite-dimensional analogue of the Kuramoto order parameter \cite{Acebron2005,PhysRevResearch.2.033025, DeVille2019}

The distinction between \(R_{12}\), and the individual quantities \(R_1\) and \(R_2\) is a central idea in this paper. The latter describes how strongly each spin is phase-localized relative to an external drive, whereas \(R_{12}\) describes how strongly the spins are synchronized with each other.  Two spins can have weak individual phase localization but possess a sharply defined relative phase. Conversely, strong entrainment of both spins by the same external drive does not, by itself, distinguish direct spin–spin synchronization from their independent locking to the common drive. A bipartite synchronization gap $ \Delta_R^{(2)} = R_1+R_2-R_{12}$ can therefore be identified, which compares the relative-phase localization of the two-spin state with the phase localization contained in the individual reduced states. A negative $\Delta_R^{(2)}$ indicates that the relative phase is more strongly localized than can be accounted for by the individual spin coherences, i.e., the mutual synchronization is not reducible to the separate entrainment of the spins by the external field.

The distinction between local entrainment and mutual synchronization becomes even more important in multipartite networks. Synchronization in multipartite systems may arise not only from pairwise interactions or individual spin contributions but also from global phase locking involving all subsystems simultaneously. This is conceptually reminiscent of genuine multipartite entanglement, which is generally not reducible to pairwise entanglement. This question of multipartite synchronization is particularly prominent in {\it finite-dimensional systems} (e.g., spin-1/2 systems) interacting with each other, either directly or indirectly through an intermediate subsystem~\cite{Ghildiyal2025, Ghildiyal2026}. In a three-spin system, we can construct a suitable tripartite synchronization gap that identifies how the collective synchronization is distributed across the network, i.e., whether it is bounded by synchronization measures obtained from reduced one- and two-spin states or not. The finite-dimensional Hilbert space and analytical tractability of the Lindblad dynamics make spin systems a natural testbed for studying synchronization in both transient and steady-state regimes~\cite{Roulet2018, Karpat2019Quantum}.

In this regard, we analyze driven-dissipative cavity-mediated two- and three-spin networks in the dispersive adiabatic regime. The external driving fields sustain the nonequilibrium dynamics, while photon hopping between the cavities mediates effective interactions among the spins. In the dispersive regime, the cavity modes can be adiabatically eliminated, producing effective two- and three-spin dynamics governed by Lindblad master equations. This architecture allows us to distinguish the phase localization of each spin relative to the drive from the relative-phase localization generated between the spins. For two spins, we find that the bipartite synchronization gap is negative in the parameter regime studied. For a triangular topology of three spins, bipartite synchronization dominates over the collective one in the relevant parameter regime. 

We note that quantum Fisher information (QFI) provides an operationally grounded measure of synchronization, linking phase locking to parameter estimation precision via the Cramér-Rao bound~\cite{Vaidya2024, Shen2023}. Experimentally, synchronization has been demonstrated across cold atomic ensembles~\cite{Laskar2020}, trapped ions~\cite{Zhang2023}, NMR systems~\cite{PhysRevA.105.062206}, and superconducting circuit QED~\cite{Tao2025}, confirming phase locking down to the single-spin level and establishing noise-induced synchronization through common dissipative baths~\cite{Tyagi2024}.

{\bf Measure of mutual spin-spin synchronization:}

Let us first consider two spins characterized by phase variables
$\phi_1,\phi_2\in[0,2\pi)$ (in their respective phase space or the Bloch sphere in the case of a single spin) with a joint probability density
$P_{12}(\phi_1,\phi_2)$. The corresponding marginal phase distributions are
defined as
\begin{equation}
\begin{aligned}
P_j(\phi_j)&=\int_0^{2\pi}P_{12}(\phi_1,\phi_2)\, d\phi_k\;,\;\;\; j\ne k\in \{1,2\}
\end{aligned}
\end{equation}
which satisfy the normalization condition
$\int_0^{2\pi} P_j(\phi_j)\, d\phi_j = 1,\; j\in\{1,2\}$. We define the single-spin
phase synchronization measure through the first moment of the phase
distribution,
\begin{equation}
R_j=\left|\int_0^{2\pi}P_j(\phi_j)\,e^{i\phi_j}\, d\phi_j\right|\;,\;\;\;j\in\{1,2\}.
\label{eq:R_def1}
\end{equation}
This quantity measures the degree of individual phase localization of the $j$th spin. It vanishes for a uniform
distribution $P_j(\phi_j)=1/(2\pi)$ and becomes finite when phase locking is
present. We next define a bipartite synchronization measure 
\begin{equation}
R_{12}=\left|\int_{-\pi}^{\pi}P_{12}(\phi_-^{(12)})\,e^{i\phi_-^{(12)}}\, d\phi_-^{(12)} \right|\;,
\label{2qbit}
\end{equation}
where $P_{12}(\phi_-^{(12)})$ denotes the joint phase distribution projected onto the space of relative phase $\phi_-^{(12)} = (\phi_1-\phi_2)/2$ of the two subsystems. This measure characterizes the localization of this relative phase $\phi_-^{(12)}$. In particular, $R_{12}=0$ for a uniform relative-phase distribution, whereas a distribution localized around a preferred phase difference yields a finite value. The modulus in Eq.~\eqref{2qbit} removes the dependence on the choice of the global phase reference, making $R_{12}$ a measure of relative-phase localization. The larger $R_{12}$ is, the stronger the phase localization is. This definition is analogous to the order parameter commonly used in classical synchronization theory of Kuramoto~\cite{Kuramoto2003,Acebron2005}. Note that $R_j$ and $R_{12}$ lie in the range $[0,1]$. Therefore, for a symmetric configuration, $R_1=R_2<0.5$ defines a regime when the mutual synchronization is expected to dominate the individual phase locking. More importantly, when the two spins do not have individual phase localization, i.e., when $R_1=R_2=0$, one can still have nonzero $R_{12}$ indicating strong relative phase synchronization. 

The extension to tripartite systems follows analogous reasoning. We consider three spins characterized by phase variables
$\phi_j\in[0,2\pi)$ ($j\in 1,2,3$) with normalized joint phase distribution
$P_{123}(\phi_1,\phi_2,\phi_3)$.
The corresponding bipartite marginals are
\begin{equation}
P_{ij}(\phi_i,\phi_j)=\int_0^{2\pi}P_{123}(\phi_1,\phi_2,\phi_3)\,d\phi_k,
\end{equation}
while the single-party marginal distribution is
\begin{equation}
P_i(\phi_i)=\int_0^{2\pi}\!\!\int_0^{2\pi}P_{123}(\phi_1,\phi_2,\phi_3)\,d\phi_j\,d\phi_k,
\end{equation}
where $i,j,k\in 1,2,3$ and $i\ne j\ne k$. We project the tripartite distribution onto the relative phase variables $\phi_-^{(ij)}=(\phi_i-\phi_j)/2$:
\begin{equation}
P_{123}(\phi_-^{(12)},\phi_-^{(23)})
=
\int_0^{2\pi}
P_{123}
(2\phi_-^{(12)}+\phi_2,\phi_2,\phi_2-2\phi_-^{(23)})
\,d\phi_2.
\label{eq:proj}
\end{equation}
Then, the tripartite synchronization measure can be obtained from the above distribution as
\begin{align}
{R}_{123}&=\left|\int_{-\pi}^{\pi}d\Phi_- e^{i\Phi_-}\int_{\mathcal{D}_{+}}d\Phi_+ P_{123}(\phi_-^{(12)},\phi_-^{(23)})\right|\;,
\end{align}
where  $\Phi_{\pm} = (\phi_-^{(12)}\pm \phi_-^{(23)})/2$ and $\mathcal{D}_+$ denotes relevant limit of $\Phi_+$. Note that $\Phi_- = (\phi_1-2\phi_2+\phi_3)/4$ defines a joint phase function, capturing the essence of three-body phase locking. This is in analogy of the phase relation of three coupled pendulums prepared in one of the normal modes. 
We define a tripartite synchronization gap $\Delta_R^{(3)}=
R_{12}(t)+R_{23}(t)-R_2(t)-R_{123}(t)$ to indicate the interplay of synchronization of single-spin, two-spin, and three-spin systems. Different partitioning among the spins can elucidate the rich synchronization architecture in a tripartite system. In the presence of inherent symmetry in the spin-spin coupling in our model, this metric lies in the range [-1,1]. In case of complete phase disorder, $R_{12}$, $R_{23}$, $R_{2}$, and $R_{123}$ can all be simultaneously zero. On the other hand, in the case of complete synchronization, the bipartite synchronization exactly conforms with the tripartite one and therefore, $R_{12}, R_{23}, R_{123} =1$, so that $\Delta_R^{(3)}$ becomes unity. This happens when $R_2=0$, which takes the global phase into account. Most interestingly, when the pairwise order parameters are small or zero and the triadic parameter $R_{123}\lesssim 1$, there can exist higher-order synchronization \cite{Zhai2026Twisted} without pairwise correlation, referring to a negative synchronization gap: $0>\Delta_R^{(3)}\ge -1$. For three spins, this is related to an emergent quantum splay state \cite{PhysRevLett.134.033801,Satoori2023Quantum}. Such tripartite synchronization is associated with a quantum correlation that cannot be accounted for by bipartition.

In the following, we investigate the behavior of spin-spin mutual synchronization and the synchronization gaps in two cavity QED setups: one with two spins and the other with three spins (as in a quantum Rabi model). We replace the phase distribution with the quasi-probability distribution. The phase distributions $P_j(\phi_j)$ are obtained through angular marginalization of the corresponding Husimi Q functions. The synchronization measures, therefore, correspond directly to the first angular moments of these distributions.

\begin{figure}[ht]
\includegraphics[height=4cm,width=8 cm]{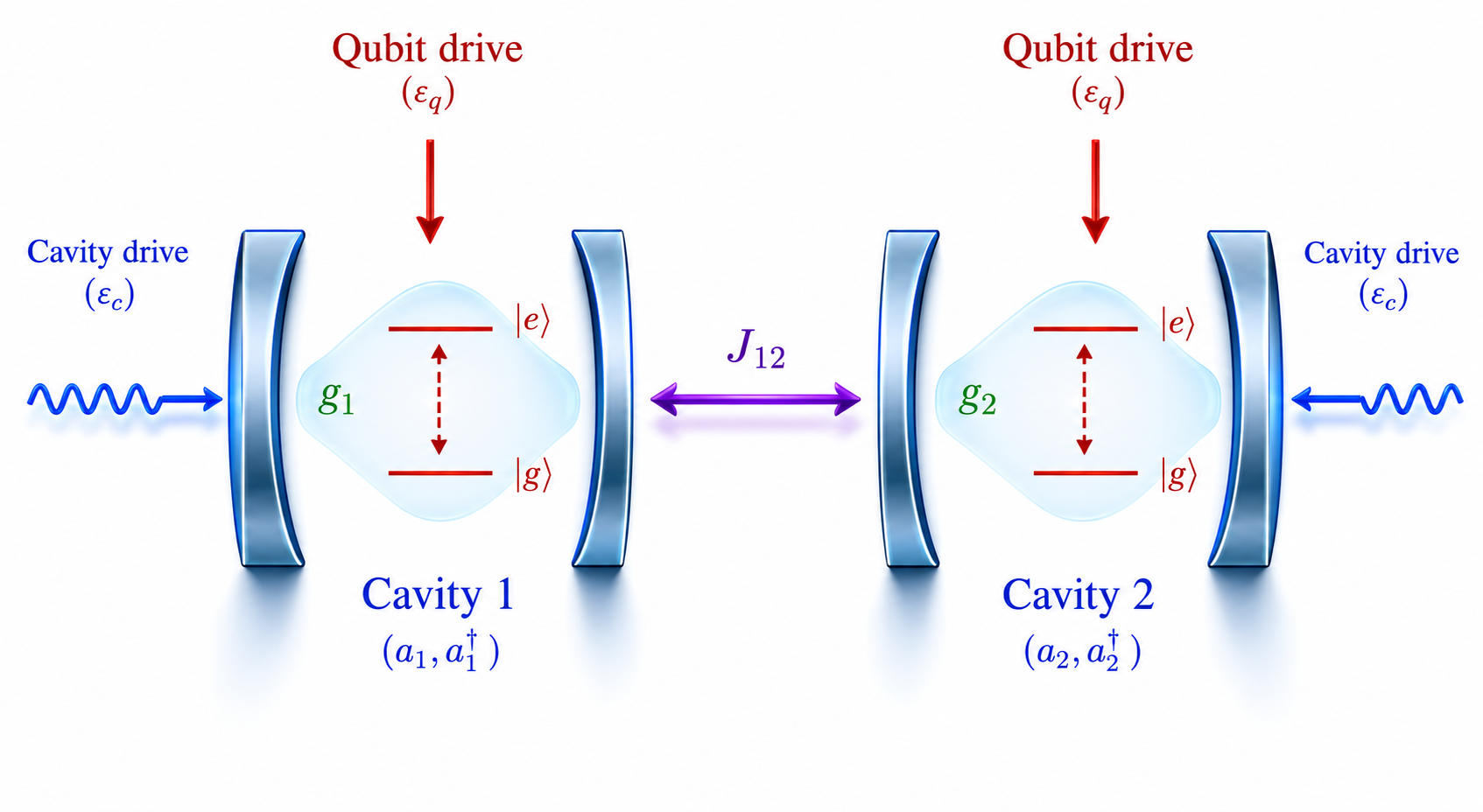}
\caption{Schematic representation of two spins inside a coupled-cavity system.}
\label{fig:1}
\end{figure}

{\bf Cavity-coupled two-spin systems:}
We consider two spatially separated spins, each coupled to a respective local cavity mode via a Jaynes-Cummings-type interaction. In contrast, coherent hopping between the cavity modes mediates an effective spin-spin interaction (see Fig.~\ref{fig:1}). The relevant Hamiltonian under the rotating wave approximation (RWA) is given by 
\begin{equation}
\begin{aligned}
H_{\mathrm{RWA}} &=\sum_{j=1}^{n}\Big[\Delta_c a_j^{\dagger} a_j+ \frac{1}{2}\Delta_j\sigma_z^{(j)}+ g_j \left(a_j^{\dagger} \sigma_-^{(j)} + \sigma_+^{(j)} a_j \right)\\
&+ \frac{1}{2}\varepsilon_{q_j}\sigma_x^{(j)}+ \varepsilon_c (a_j + a_j^{\dagger})\Big]
+ \sum_{\substack{i=1 \\ i \neq j}}^n J_{ij}\left(a_i^{\dagger} a_j + a_j^{\dagger} a_i \right),
\label{hamilrwa}
\end{aligned}
\end{equation}
with $n=2$.
Here $a_j$ ($a_j^\dagger$)is the annihilation (creation) operator of the $j$th cavity,  $\sigma_z^{(j)},\sigma_\pm^{(j)}$ denote Pauli operators for the $j$th spin in the basis $\{|g\rangle,|e\rangle\}$. The $g_j$ is the spin–cavity coupling strength, $J_{ij}$ is the photon hopping amplitude between cavities, and $\varepsilon_c$ and $\varepsilon_{q_j}$ specify coherent drives of the cavity and $j$th spin, with respective detunings $\Delta_c$ and $\Delta_j$.

In the large cavity decay rate limit, we adiabatically eliminate the cavity mode to obtain an effective two-spin Hamiltonian in the dispersive regime ($g_j< \Delta_j, \gamma_j<\kappa <\Delta_c$; $\kappa$ is the cavity decay rate) [see \cite{SM}].
In this adiabatic regime, we solve the Lindblad master equation for the two-spin density matrix $\rho(t)$ in the presence of local relaxation and pure local dephasing for each spin. The collective Husimi $Q$ distribution can thus be written as 
$Q_{\mathrm{12}}(\theta,\phi_1,\phi_2,t)=\frac{1}{(2\pi)^2}\langle\psi_{12}|\rho(t)|\psi_{12}\rangle,$
where $|\psi_{12}\rangle = |\theta,\phi_1\rangle\otimes |\theta,\phi_2\rangle$ is a two-spin coherent state 
and $|\theta_j,\phi_j\rangle =\cos(\theta_j/2)|g\rangle+ e^{i\phi_j}\sin(\theta_j/2)|e\rangle$ is the spin-$\tfrac{1}{2}$ coherent state with $\theta_j \in [0,\pi]$ and $\phi_j \in [-\pi,\pi)$, denoting the polar and azimuthal coordinates on the Bloch sphere, respectively.

The two-spin quantum phase synchronization measure $R_{12}(t)$ [Eq. (\ref{2qbit})] can then be obtained in terms of the following normalized phase distribution (for $\theta_j=\theta$):
\begin{equation}
P_{12}(\phi_-^{(12)},t)= \int_{0}^{\pi}d\theta\;Q_{\mathrm{12}}(\theta,\phi_-^{(12)},t)\sin\theta,
\end{equation}
where $\phi_-^{(12)}\in [-\pi,\pi]$ and 
\begin{equation}
    Q_{12}(\theta,\phi_-^{12)},t)=2\int_{|\phi_-^{(12)}|}^{2\pi-|\phi_-^{(12)}|} d\phi_+^{(12)}\; Q_{12}(\theta,\phi_1,\phi_2,t)\;.
\end{equation}
The factor of 2 appears as the Jacobian for a coordinate transformation $\phi_{1,2}\rightarrow \phi_\pm^{(12)}$. 
 Similar phase-space measures based on the Husimi-$Q$ representation have also been employed in studies of quantum synchronization ~\cite{Roulet2018,PhysRevLett.121.053601}. 

Similarly, the single-spin synchronization $R_j(t)$ can be obtained using (\ref{eq:R_def1}) in terms of the phase-distribution $P_j(\phi_j)$ given by  $
P_j(\phi_j)=\int_0^\pi Q_j(\theta_j,\phi_j)\sin\theta_j\,d\theta_j$,
which encodes the azimuthal statistics of each spin independent of its polar 
orientation. Here, $Q_j=\frac{1}{2\pi}
\langle\theta_j,\phi_j|\rho_j|\theta_j,\phi_j\rangle$ is the normalized projection of the reduced density matrix $\rho_j(t)$ of the $j$th spin on the
spin-$\tfrac{1}{2}$ coherent states $
|\theta_j,\phi_j\rangle$.
 Together, 
$R_{12}$, $R_1$, and $R_2$ provide a complete and hierarchical characterization 
of phase coherence at both the collective and individual-subsystem levels.

\begin{figure}[ht]

     \subfloat[\label{2a}]{%
		\includegraphics[height=3.5cm,width=0.6\linewidth]{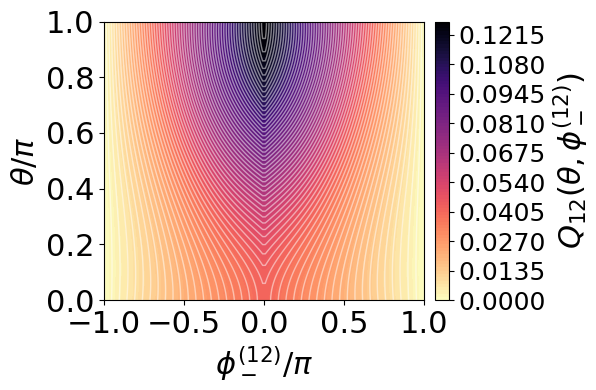}%
	 }\\
      \subfloat[\label{2b}]
      {%
		\includegraphics[height=3.5cm,width=0.5\linewidth]{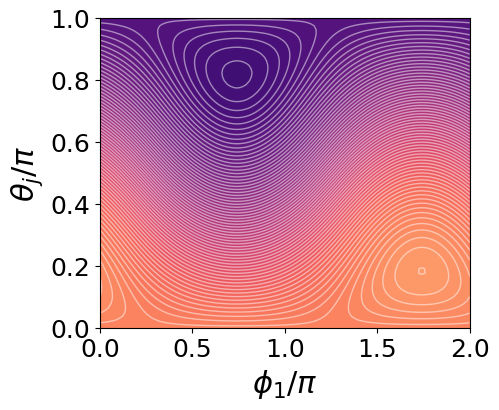}%
	 }
        \subfloat[\label{2c}]{%
		\includegraphics[height=3.5cm,width=0.55\linewidth]{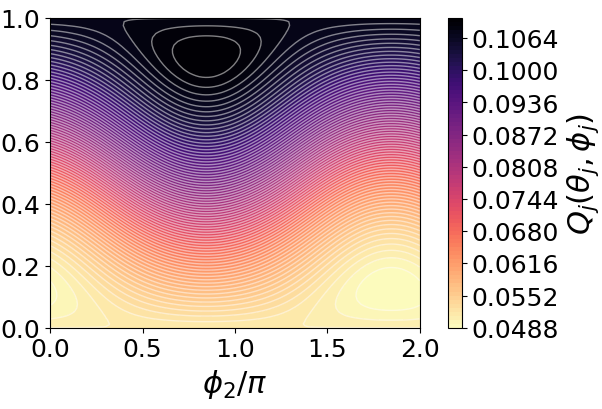}
	}
        

\caption{Density plots of steady-state Husimi-$Q$ distributions (a) $Q_{12}$, (b) $Q_1$, and (c) $Q_2$, in their respective parameter spaces.
We have chosen $\Delta_c=5$, $\Delta_{1,2}=(1,2)$, $J_{12}=1$,
$g_{1,2}=0.2$, $\varepsilon_{q,c}=1.0$, $\kappa=1$,
the relaxation rate $\gamma_{1,2}=0.1$, and
the dephasing rate $\gamma_{\phi_{1,2}}=0.5$. All parameters are normalized in terms of $J_{12}$.}
\end{figure}

\begin{figure}[ht]
	  \subfloat[\label{3a}]{%
		\includegraphics[height=3cm,width=0.46\linewidth]{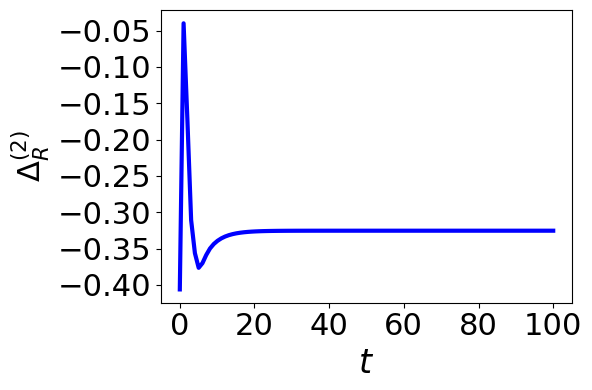}%
	}
     \subfloat[\label{3b}]{%
		\includegraphics[height=3cm,width=0.46\linewidth]{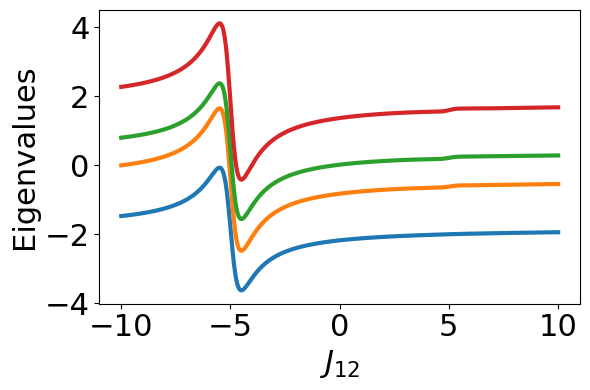}%
	}\\
    
	  \subfloat[\label{3c}]{%
	    \includegraphics[height=3.3cm,width=0.55\linewidth]{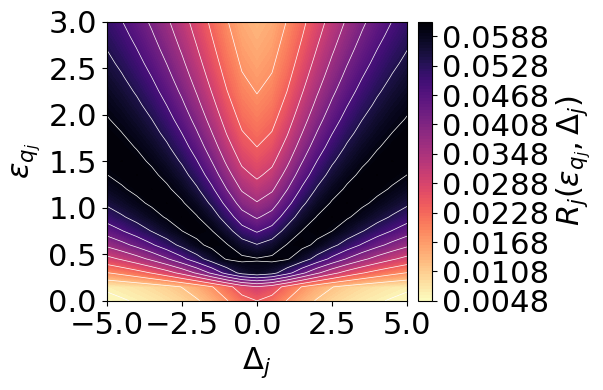}%
    }
    \subfloat[\label{3d}]{%
		\includegraphics[height=3.3cm,width=0.55\linewidth]{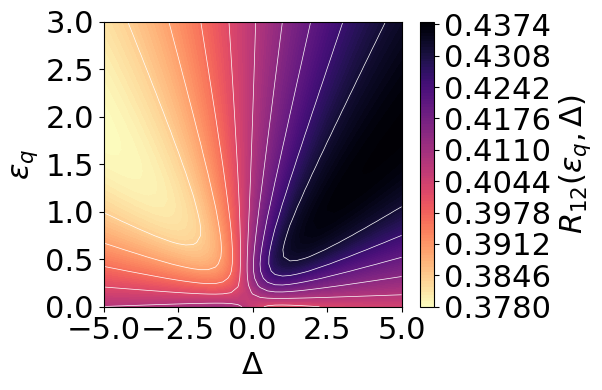}%
	}\\
        \subfloat[\label{3e}]{%
		\includegraphics[height=3.5cm,width=0.66\linewidth]{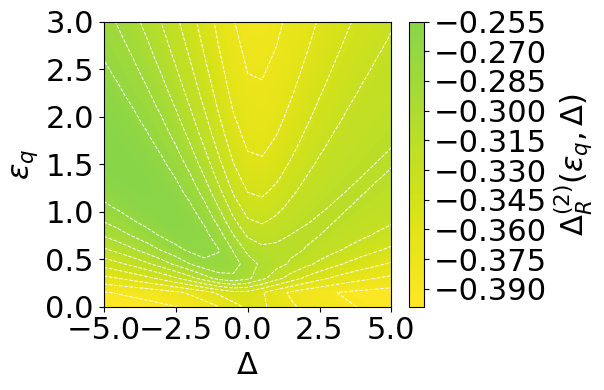}%
	}
\caption{Time evolution of (a) the synchronization gap $\Delta_R^{(2)}(t)$ for $J_{12}=1$ and (b) the effective-Hamiltonian eigenvalue spectrum versus $J_{12}$. Panels (c) and (d) show the steady-state synchronization measures $R_j(\epsilon_{q_j},\Delta_j)$ and $R_{12}(\epsilon_q,\Delta)$, respectively, and (e) shows the corresponding synchronization gap $\Delta_R^{(2)}(\epsilon_q,\Delta)$. All other parameters are as in Fig.~2, except in panels (c)-(e), in which we have chosen $\Delta_1=\Delta_2=\Delta$.}
\end{figure}

Figures~\ref{2a}-\ref{2c} show the steady-state Husimi $Q$ distributions
$Q_{12}(\theta,\phi_-^{(12)})$, $Q_1(\theta_1,\phi_1)$, and
$Q_2(\theta_2,\phi_2)$. The joint distribution (Fig.~\ref{2a}) is sharply
peaked at $\phi_-^{(12)}=0$ over a broad range of $\theta$, signaling a
well-defined relative phase between the two spins. The single-spin
distributions (Figs.~\ref{2b}, \ref{2c}), by contrast, are spread over a
wider region of phase space, each showing a broad maximum near
$\theta_j/\pi\sim0.8$--$0.9$, $\phi_j/\pi\sim0.8$--$1.0$ and a narrower
feature near $\theta_j/\pi\sim0.1$--$0.2$, $\phi_j/\pi\sim1.7$--$2.0$. Each spin is thus weakly localized in phase, while their joint state is
tightly localized in the relative phase $\phi_-^{(12)}$. This stronger localization is a direct signature
of cavity-mediated phase correlations that cannot be inferred from either
reduced state alone.

This asymmetry is quantified by the synchronization gap
$\Delta_R^{(2)}(t)=R_1(t)+R_2(t)-R_{12}(t)$ [see Fig.~\ref{3a}]. After a transient variation, it saturates to a negative value of $-0.32$ at the steady state $t\gtrsim20/J_{12}$. Since this negative value implies 
$R_{12}>R_1+R_2$ at steady state, the joint state is said to be more
strongly synchronized than each spin is to its respective drive throughout the evolution and not
merely during the transient. This superadditivity is the direct dynamical
counterpart of the $\phi_-^{(12)}$ localization in $Q_{12}$. The cavity-mediated
coupling between the spins correlates their phases in respective Bloch spheres. Thus, the nonzero $R_{12}$ carries relative-phase correlation, which cannot be extracted from $R_1$ and $R_2$ alone. The $R_j$'s register
only the local phase spread of each reduced state. The steady negative value
of $\Delta_R^{(2)}$ confirms that this collective enhancement in phase correlation is an intrinsic
property of the coupled steady state, not a transient artifact.

In Figs.~\ref{3b}-\ref{3e}, we show how this enhancement manifests itself across a large range of coupling and
detuning parameters. In Fig.~\ref{3b}, we plot the eigenvalues of the coupled
two-spin Hamiltonian [see Eq. (3),\cite{SM}] against $J_{12}$. It is clear that away from $J_{12}\simeq-5$ the four
levels vary smoothly toward well-separated asymptotes as $|J_{12}|\to\infty$. This limit refers to an effective weak coupling between the spins such that each spin retains its nearly independent character. Near
$J_{12}\simeq-5$, the two intermediate energy levels approach each other while the outer
pair repels. This regime marks an avoided crossing that leads to the strongest hybridization between two spins
and, correspondingly, the coupling regime where phase correlations are
strongest.

In Fig.~\ref{3c} we map the single-spin measure $R_j(\varepsilon_{q_j},\Delta_j)$ over the drive-detuning plane. The single-spin synchronization exhibits an Arnold-tongue structure: $R_j$ is
maximum along $\Delta_j=0$ for weak driving and is reduced significantly as the driving strength $\varepsilon_{q_j}$ increases beyond 1.5. Larger $R_j$ corresponds to tighter phase localization, so each spin
synchronizes best under resonant, weak driving; $R_j$ remains below
$\sim0.06$ in this region.
Unlike the usual Arnold tongue reported in \cite{koppenhofer2019optimal,PhysRevLett.121.053601,PhysRevA.101.062104,Roulet2018}, where stronger synchronization is found within the tongue region, we found the opposite. In our case, the tongue region refers to the stronger desynchronization for larger driving. Synchronization occurs outside the tongue region and can be maintained for a larger range of nonzero $\Delta_j$ and $\varepsilon_{q_j}$. Similar results can also be found in single-spin phase localization in a non-Markovian bath \cite{ali2024detecting}. The desynchronization region widens as the drive strength increases. In a similar way, we plot in Fig.~\ref{3d} $R_{12}(\varepsilon_q,\Delta)$ over the same plane.
Its values, $0.38-0.44$, are an order of magnitude larger than $R_j$ everywhere and follow the same Arnold-tongue-like dependence on
$\Delta_j=\Delta$ and $\varepsilon_{q_j}=\varepsilon_q$, with a mild asymmetry favoring $\Delta>0$ at
larger $\varepsilon_q$. Clearly, the
inequality $R_{12}>R_1+R_2$ as established in Fig.~\ref{3a} holds throughout
parameter space rather than at an isolated operating point. Thus, the cavity-mediated enhancement of joint synchronization over the sum of individual contributions is a structural feature of the coupled system, not an artifact of the specific parameters used for $\Delta_R^{(2)}(\varepsilon_q,\Delta)$ as shown in Fig.~\ref{3e}. Note that the value of $\Delta_R^{(2)}$ becomes more negative as the resonant driving $\varepsilon_q$ is increased. This is because the single-spin synchronization deteriorates for larger driving at resonance while the mutual synchronization remains approximately unaltered.

\begin{figure}[ht]
\includegraphics[height=6cm,width=8.5cm]{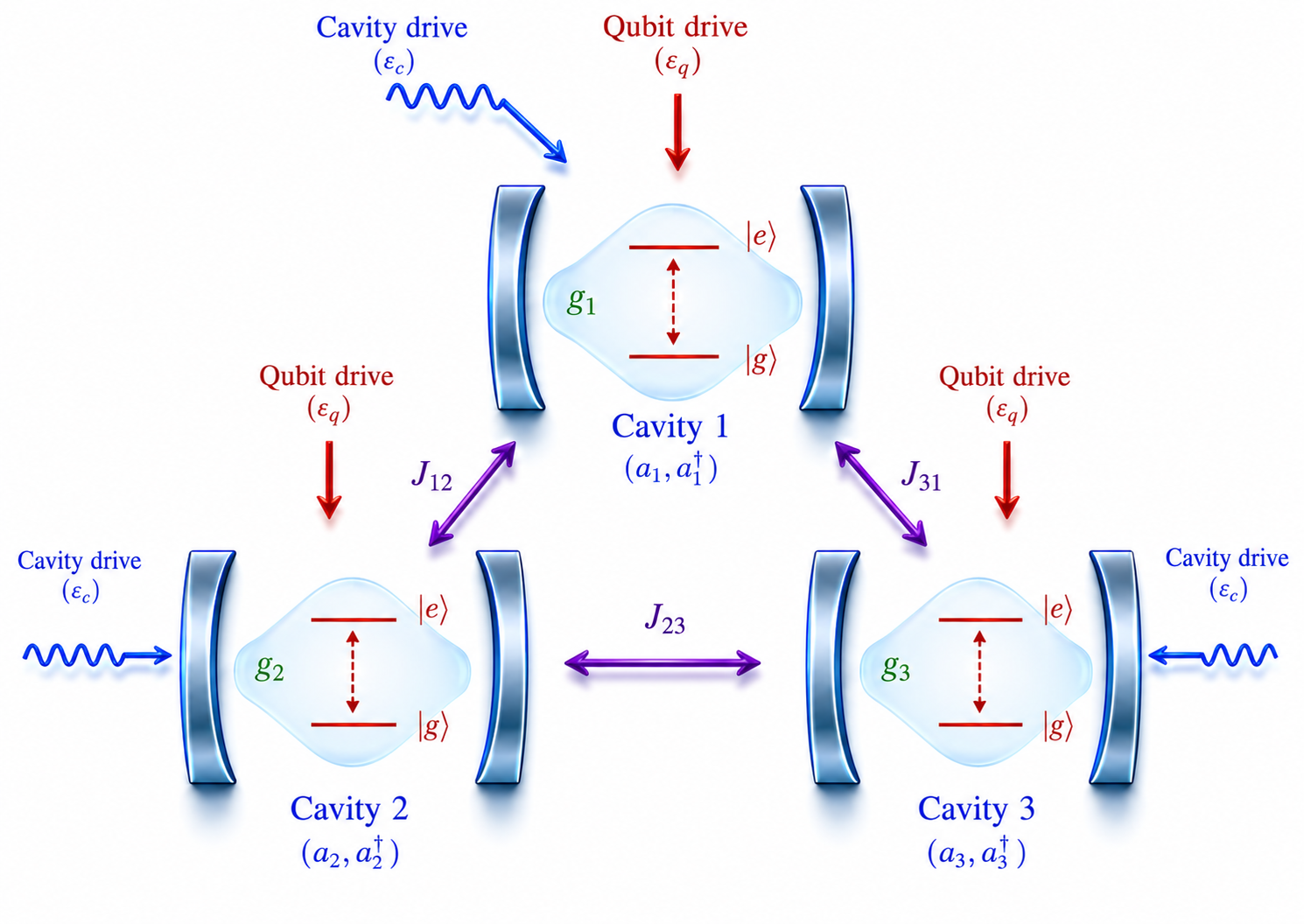}
    \caption{Schematic representation of three spins inside a triangular cavity-coupled system.}
    \label{fig:4}
\end{figure}

{\bf Three-spin quantum Rabi model:}
We next investigate mutual synchronization among three spins, each coupled to local cavities arranged in a triangular configuration, while these cavities are mutually coupled through photon hopping (Fig.~\ref{fig:4}). We adiabatically eliminate the cavity modes from the corresponding Hamiltonian (\ref{hamilrwa}) for $n=3$. The effective Hamiltonian thus obtained in the dispersive limit [see Eq. (9), \cite{SM}] describes the dynamics of a three-coupled-spin system. We solve the Lindblad master equation for the tripartite density matrix $\rho_{123}(t)$, with this Hamiltonian in the presence of local relaxation and local dephasing.

To obtain the synchronization measure, 
we next compute the Husimi Q-function $Q_{123}(\theta,\{\phi_j\},t)=(1/2\pi)^3\langle\psi_{123}|\rho_{123}(t)|\psi_{123}\rangle,$ where $|\psi_{123}\rangle = \otimes_{j=1}^3|\theta,\phi_j\rangle$ is the spin-$\tfrac{3}{2}$ coherent state.
The tripartite synchronization is therefore characterized by the first azimuthal moment of the Husimi distribution as
\begin{equation}
R_{123}(t)=\left|
\int_{0}^{\pi} d\theta\,\sin\theta
\int_{-\pi}^{\pi} d\Phi_-\,
Q_{123}(\theta,\Phi_-,t)\, e^{i\Phi_-}
\right|,
\end{equation}
which quantifies the degree of phase localization in the collective
distribution. Here $Q_{123}(\theta,\Phi_-,t)$ is the Husimi Q-function averaged over $\phi_2$ and $\Phi_+$ over suitable range [see \cite{SM}, for details]. For a phase distribution uniform in $\Phi_-$,
$R_{123}(t)$ vanishes, whereas finite values indicate the emergence of a
preferred phase direction. As was done in the two-spin case, the 
synchronization measures for reduced subsystems are obtained from the 
corresponding reduced density matrices. 

\begin{figure}[ht]

     \subfloat[\label{5a}]{%
		\includegraphics[height=3cm,width=0.5\linewidth]{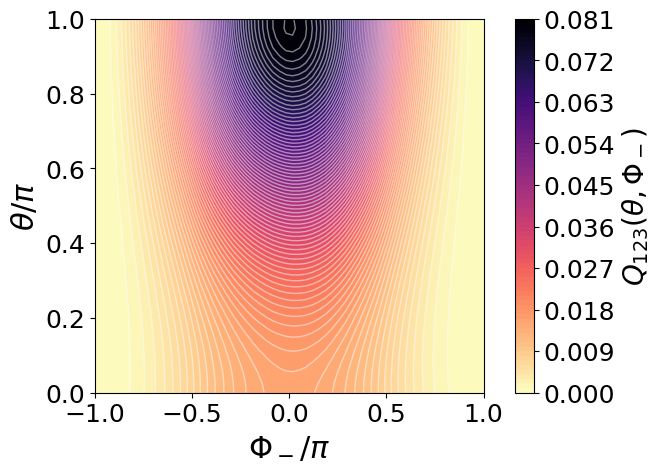}%
	 }\\
      \subfloat[\label{5b}]{%
		\includegraphics[height=2.5cm,width=0.34\linewidth]{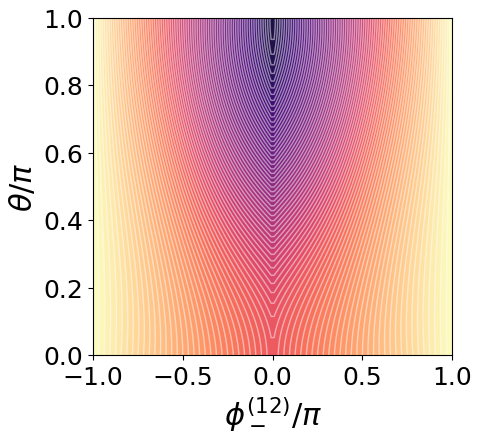}%
	 }
        \subfloat[\label{5c}]{%
		\includegraphics[height=2.5cm,width=0.32\linewidth]{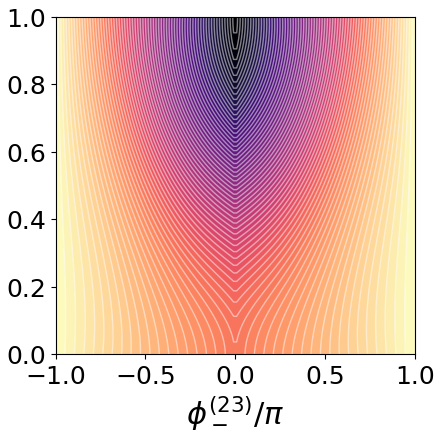}
	}
	  \subfloat[\label{5d}]{%
		\includegraphics[height=2.5cm,width=0.40\linewidth]{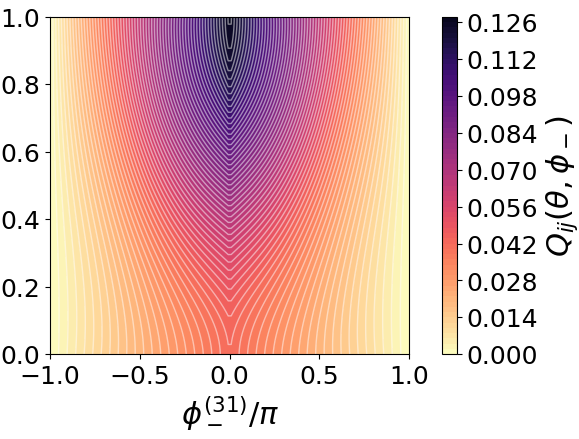}%
	}\\

    \subfloat[\label{5e}]{%
		\includegraphics[height=2.5cm,width=0.34\linewidth]{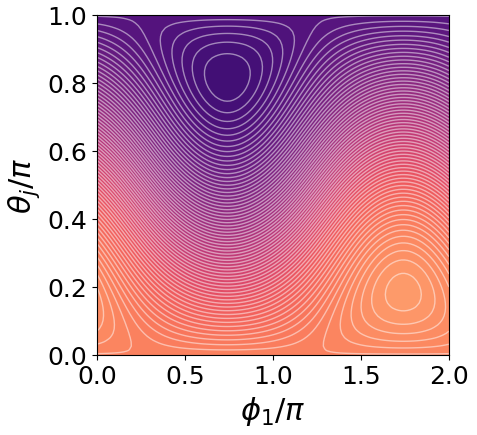}%
	 }
        \subfloat[\label{5f}]{%
		\includegraphics[height=2.5cm,width=0.32\linewidth]{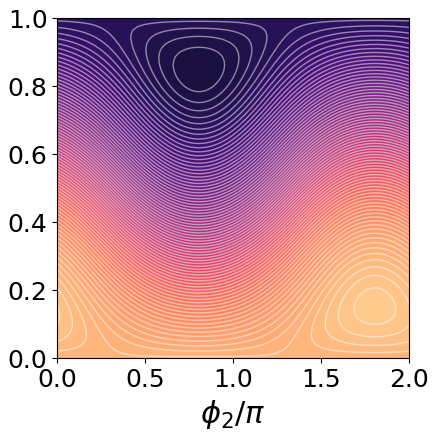}
	}
	  \subfloat[\label{5g}]{%
		\includegraphics[height=2.5cm,width=0.40\linewidth]{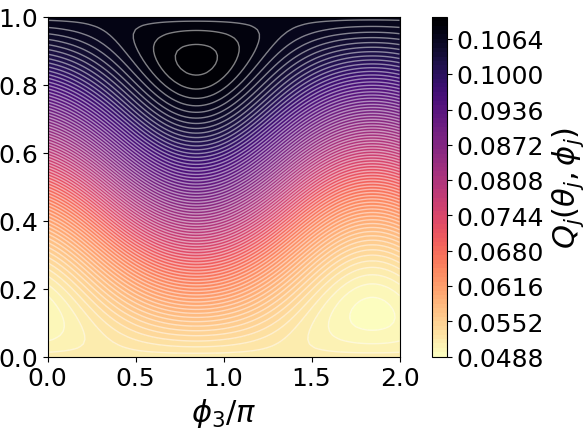}%
	}\\

\caption{Density plots of the Husimi-$Q$ distributions (a) $Q_{123}$, (b) $Q_{12}$, (c) $Q_{23}$, (d) $Q_{31}$, (e) $Q_1$, (f) $Q_2$, and (g) $Q_3$ in their respective phase space.
We have chosen $\Delta_c=5$, $\Delta_{1,2,3}=(1,1.5,2)$,
$J_{12}=J_{23}=J_{31}=1$, $g_{1,2,3}=0.2$,
$\varepsilon_{q,c}=1.0$, $\kappa=1$, $\gamma_{1,2,3}=0.1$, and
$\gamma_{\phi_{1,2,3}}=0.5$.}
\end{figure}

\begin{figure}[ht]
     \subfloat[\label{6a}]{%
		\includegraphics[height=3cm,width=0.46\linewidth]{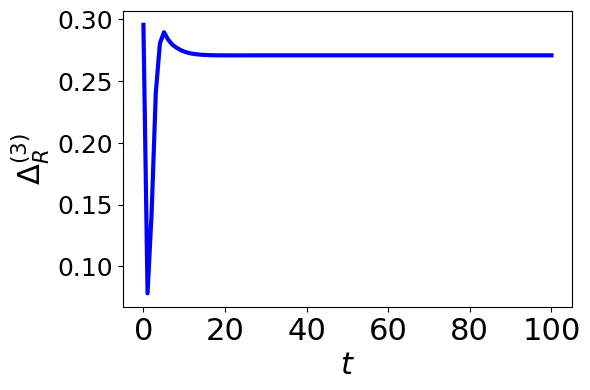}%
	}\\
	  \subfloat[\label{6b}]{%
	    \includegraphics[height=3.2cm,width=0.52\linewidth]{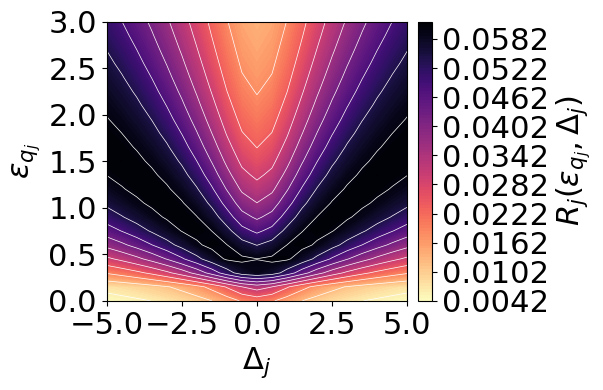}%
    }
    \subfloat[\label{6c}]{%
	    \includegraphics[height=3.2cm,width=0.52\linewidth]{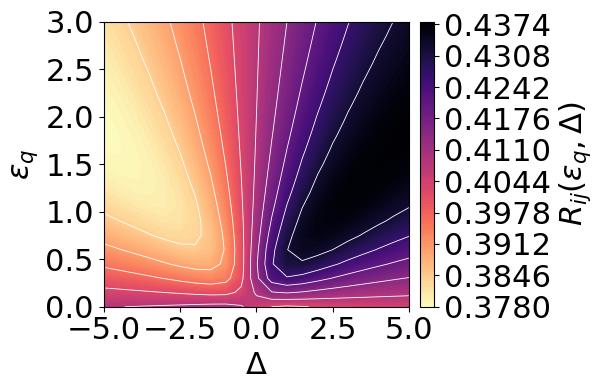}%
    }\\
    \subfloat[\label{6d}]{%
	    \includegraphics[height=3.2cm,width=0.52\linewidth]{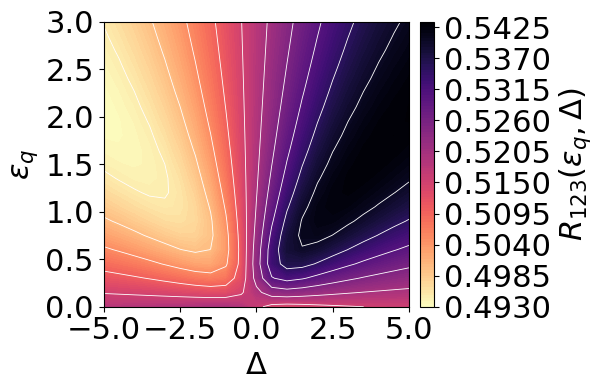}%
    }
    \subfloat[\label{6e}]{%
	    \includegraphics[height=3.2cm,width=0.52\linewidth]{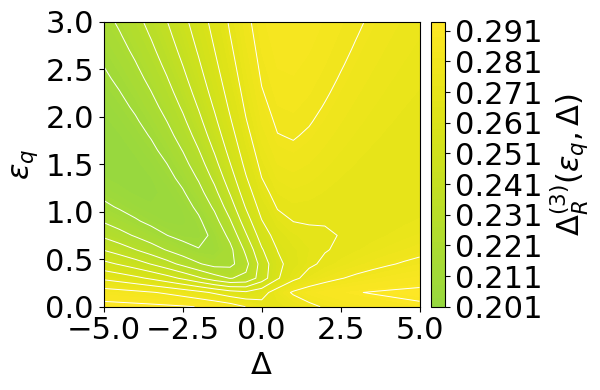}%
    }
\caption{Time evolution of the three-spin synchronization gap $\Delta_R^{(3)}(t)$ in (a). Panels (b)-(d) show the steady-state synchronization measures $R_j$, $R_{ij}$, and $R_{123}$, respectively, as functions of $(\epsilon_{q_j},\Delta_j)$ or $(\epsilon_q,\Delta)$, while (e) shows the corresponding synchronization gap $\Delta_R^{(3)}$. All other parameters are as in Fig.~5, except in panels (b)-(e), in which we have chosen $\Delta_1=\Delta_2=\Delta$.}
\end{figure}

The steady-state Husimi-$Q$ distributions for the three-spin system are
shown in Fig.~\ref{5a} and Figs.~\ref{5b}-\ref{5g}. The collective
distribution $Q_{123}(\theta,\Phi_-)$ is shown in Fig.~\ref{5a}, the
pairwise reduced distributions $Q_{ij}(\theta,\phi_-^{(ij)})$ for
$(ij)=(12),(23),(31)$ in Figs.~\ref{5b}--\ref{5d}, and the single-spin
distributions $Q_j(\theta_j,\phi_j)$ in Figs.~\ref{5e}-\ref{5g}. The
collective distribution is localized around $\Phi_-=0$, and the three
pairwise distributions are likewise concentrated around
$\phi_-^{(ij)}=0$. The larger peak amplitudes of the pairwise
distributions indicate stronger localization of the corresponding
relative phases. In contrast, the single-spin distributions are broader
in the azimuthal direction and show a diffuse maximum around
$\theta_j/\pi\simeq0.8$-$0.9$. The phase-space localization is
therefore more pronounced in the relative-phase variables than in the
individual spin phases, consistent with correlated phase dynamics
generated by the coupling of the spins to the cavity modes.

To quantify the relation between the collective and the reduced
synchronization measures, we define the three-spin synchronization
gap as $\Delta_R^{(3)}(t)=R_{12}(t)+R_{23}(t)-R_{2}(t)-R_{123}(t)$, in analogy to the strong subadditivity inequality of von Neumann entropy. We must emphasize that the metric for quantum synchronization is based on relative phase distribution, while the entropy is a function of both occupation probabilities and the coherences. Therefore, these two markers for correlations must be understood as fundamentally different. 
Its time evolution is shown in Fig.~\ref{6a}.  
We find that after a transient phase, this synchronization gap settles to a
positive stationary value, $\Delta_R^{(3)}(t\to\infty)\simeq0.27$. In
the steady state we therefore have $R_{123}\leq R_{12}+R_{23}-R_{2}$,
which implies that the collective synchronization measure is bounded by the
pairwise measures reduced by the central-spin contribution. Figures~\ref{6b}-\ref{6d} show the steady-state values of $R_j$ as functions of $(\epsilon_{q_j},\Delta_j)$ and 
$R_{ij}$ and $R_{123}$ as functions of 
$(\epsilon_{q_j}=\epsilon_q,\Delta_j=\Delta)$. Clearly, the single-spin measure $R_j$ exhibits an Arnold tongue-like structure. Outside the central region around $\Delta_j=0$, each spin exhibits better synchronization. The region of desynchronization widens with the driving strength. 
The pairwise and
collective measures remain much larger than the single-spin measures in the same parameter range. More importantly, both measures exhibit an asymmetric Arnold tongue: they remain larger in the region $\Delta > 0$ than in $\Delta < 0$. Mutual spin synchronization can thus be controlled by changing the sign of detuning. 
The corresponding synchronization gap is plotted in Fig.~\ref{6e}. Clearly, it remains
positive over the entire parameter window and takes its smallest values
in the region of negative $\Delta$ and large $\epsilon_q$. The
inequality $R_{123}\leq R_{12}+R_{23}-R_{2}$ is thus not tied to one
choice of $\epsilon_q$ and $\Delta$, but is robust against a large change in the driving parameters. The positive value of $\Delta_R^{(3)}$ indicates the existence of tripartite synchronization which is, however, less dominant than the bipartite ones. This means that there exists bipartite synchronization that may not be accounted for by the tripartite phase correlation.

{\it Conclusion:} We have formulated measures for mutual phase synchronization defined from the first angular moments of Husimi-$Q$ distributions. The measure directly probes relative-phase localization in finite-dimensional quantum systems. We investigated driven-dissipative spins coupled through tunnel-connected cavity modes and treated the cavity dynamics in the dispersive regime using adiabatic elimination.

For two spins, the joint phase distribution develops stronger relative-phase localization than the corresponding single-spin distributions, yielding, in the long-time regime, $\Delta_R^{(2)}=R_1+R_2-R_{12}<0$. Hence, $R_{12}>R_1+R_2$, demonstrating a collective synchronization contribution generated by the cavity-mediated interaction, which single-spin contributions cannot account for. For three spins, in contrast, the steady-state synchronization gap $\Delta_R^{(3)}=R_{12}+R_{23}-R_2-R_{123}$ remains positive in the parameter regime considered, giving $R_{123}\leq R_{12}+R_{23}-R_2$. Thus, it reflects a different organization of phase correlations in the three-spin network.

These results show that the subadditivity-like properties of phase synchronization are not fixed solely by the local synchronization strengths but depend on the collective dynamics generated by the interaction topology. The sign of the synchronization gap therefore provides a compact diagnostic of nonadditive synchronization contributions. Since the construction relies only on angular moments of the Husimi-$Q$ distribution, it is applicable beyond the specific cavity-mediated architecture considered here and can be extended to larger networks, alternative coupling topologies, and spatially nonuniform dissipation, providing a framework for characterizing the interplay between phase localization and many-body correlations in driven-dissipative quantum systems.

{\it Acknowledgement:} One of us (J.G.) acknowledges the financial support provided by the Department of Science and
Technology-Innovation in Science Pursuit for Inspired Research, Government of India,
through the fellowship DST/INSPIRE Fellowship/2019/IF190615 during this work.

Details of the derivation of the effective Hamiltonian and the expression of tripartite synchronization in terms of the Husimi-Q function are provided in the
Supplemental Material~\cite{SM}.

\bibliographystyle{apsrev4-2}

\bibliography{main}

\end{document}


\title{Supplemental Material for:
``Multipartite quantum synchronization in driven-dissipative spin networks''}

\author{Jatin Ghildiyal}
\email{jatin.20phz0017@iitrpr.ac.in} 
\author{Shubhrangshu Dasgupta}
\author{Asoka Biswas}
\affiliation{Department of Physics, Indian Institute of Technology Ropar, Rupnagar, Punjab 140001, India}

\maketitle

\section{Hamiltonian of two qubits}

We investigate the synchronization dynamics in a system of two spatially separated qubits that interact indirectly through photon tunneling between two coupled cavities. Each qubit is locally coupled to its respective cavity mode via a Jaynes-Cummings interaction. In contrast, coherent hopping between the cavity modes mediates an effective qubit-qubit interaction (Fig. 1 in the main text). Such cavity-assisted coupling provides a flexible platform for exploring collective dynamics and synchronization phenomena in open quantum systems.
The Hamiltonian of the system is
\begin{equation}
\begin{aligned}
H &=
\sum_{j=1}^{2}\Big[
\omega_c a_j^\dagger a_j
+ \frac{1}{2}\omega_j\sigma_z^{(j)}
+ g_j (a_j + a_j^\dagger)(\sigma_+^{(j)} + \sigma_-^{(j)})
\\
&\quad+ \varepsilon_c \big(a_j e^{-i\omega_d t} + a_j^\dagger e^{i\omega_d t}\big)
+ \frac{1}{2}\varepsilon_{q_j}\big(\sigma_+^{(j)} e^{-i\omega_q t}
+ \sigma_-^{(j)} e^{i\omega_q t}\big)
\Big] \\
&\quad
+ \sum_{\langle i,j\rangle} J_{ij}
\left(a_i^\dagger a_j + a_j^\dagger a_i\right).
\end{aligned}
\end{equation}
Here $a_j$ ($a_j^\dagger$) annihilates (creates) a photon in the $j$th cavity of frequency $\omega_c$, and $\sigma_z^{(j)},\sigma_\pm^{(j)}$ denote Pauli operators for the $j$th qubit with transition frequency $\omega_j$. The parameter $g_j$ is the qubit–cavity coupling strength, $J_{ij}$ is the photon hopping amplitude between cavities, and $\varepsilon_c$ and $\varepsilon_{q_j}$ specify coherent drives at frequencies $\omega_d$ and $\omega_q$, respectively.
Under the rotating-wave approximation, rapidly oscillating counter-rotating contributions are neglected, yielding the effective driven Hamiltonian

\begin{equation}
\begin{aligned}
H_{\mathrm{RWA}} &=
\sum_{j=1}^{2}\Big[
\Delta_c a_j^{\dagger} a_j
+ \frac{1}{2}\Delta_j\sigma_z^{(j)}
+ g_j \left(a_j^{\dagger} \sigma_-^{(j)} + \sigma_+^{(j)} a_j \right)\\
&\quad
+ \frac{1}{2}\varepsilon_{q_j}\sigma_x^{(j)}
+ \varepsilon_c (a_j + a_j^{\dagger})
\Big]
+ \sum_{\langle i,j\rangle} J_{ij}
\left(a_i^{\dagger} a_j + a_j^{\dagger} a_i \right).
\end{aligned}
\end{equation}

In the parameter regime where the cavity modes evolve on a timescale much shorter than that of the qubits, the photonic degrees of freedom may be adiabatically eliminated. We then obtain the following effective two-qubit Hamiltonian in the dispersive regime ($\Delta_c \gg \Delta_j,g_j$):

\begin{widetext}
\begin{equation}
\begin{aligned}
H
=
\sum_{j}\frac{\Delta_j}{2}\sigma_z^{(j)}
+
\sum_{j}\frac{\epsilon_{q_j}}{2}\left(\sigma_+^{(j)}+\sigma_-^{(j)}\right)
+
\frac{\Delta_c}{|D|^2}
\left[
\left(|\alpha|^2+J_{12}^2\right)S
-2J_{12}\Delta_c\,C
\right]
+
\sum_j g_j\left(Y_j^\dagger\sigma_-^{(j)}+\mathrm{H.c.}\right)&\\
\quad
+
\frac{J_{12}}{|D|^2}
\left[
\left(|\alpha|^2+J_{12}^2\right)C
-2J_{12}\Delta_c\,S
\right]
-\epsilon_c
\left[
\frac{(J_{12}+i\alpha)}{D}\sum_j X_j
+
\mathrm{H.c.}
\right],
\label{eq:Heff-general}
\end{aligned}
\end{equation}
\end{widetext}

where the first two terms are the bare qubit energies and the direct qubit drive, and the remaining terms arise from the eliminated cavity fields: a Stark-shift-like renormalization of the qubit energies, an effective cavity-mediated qubit-qubit coupling, a residual dispersive qubit-cavity interaction, and a driven cavity contribution. Here we have introduced
\begin{equation}
\alpha=\frac{\kappa}{2}+i\Delta_c, \qquad
D=\alpha^2+J_{12}^2, \qquad
X_j=g_j\sigma_-^{(j)}+\epsilon_c,
\label{eq:defs1-general}
\end{equation}
together with the bilinear combinations
\begin{equation}
S=\sum_j X_j^\dagger X_j, \qquad
C=\sum_j X_j^\dagger X_{\bar j},
\label{eq:defs2-general}
\end{equation}
and the effective coupling amplitude
\begin{equation}
Y_j=-\frac{1}{D}\left(i\alpha X_j+J_{12}X_{\bar j}\right),
\label{eq:defs3-general}
\end{equation}
where $j\in\{1,2\}$ and $\bar{j}$ denotes the complementary index ($\bar{1}=2,\ \bar{2}=1$).

Note that this Hamiltonian (\ref{eq:Heff-general}) is of Kuramoto-type (ref.~[8] in main text), in which the spins interact with a strength proportional to $\cos(\phi_1-\phi_2)$, where $\phi_{1,2}$ are their respective azimuthal coordinates. However, the primary difference lies in the local driving of each spin and the local decay, as described below by Lindbladian formalism. The decay tends to reduce the length of the Bloch vector, while in the Kuramoto model, the oscillator amplitude is maintained at unity.

With the cavity modes eliminated, the open-system dynamics of the two qubits is governed by the Lindblad master equation
\begin{equation}
\frac{d\rho}{dt}
=
-i\left[H,\rho\right]
+
\sum_{j=1}^{2}
\gamma_j\mathcal{L}\!\left[\sigma_-^{(j)}\right]\rho
+
\sum_{j=1}^{2}
\gamma_{\phi j}
\mathcal{L}\!\left[\sigma_z^{(j)}\right]\rho ,
\label{eq:master}
\end{equation}
where $\gamma_j$ and $\gamma_{\phi j}$ denote the spontaneous decay and pure dephasing rates of qubit $j$, respectively, and the dissipator takes its standard Lindblad form
\begin{equation}
\mathcal{L}[M]\rho
=
M\rho M^\dagger
-\frac{1}{2}
\left\{
M^\dagger M,\rho
\right\}.
\label{eq:dissipator}
\end{equation}

\section{Hamiltonian of three qubits}

Assuming that the cavity modes can be adiabatically eliminated, i.e., that the cavity detuning, decay, and inter-cavity hopping occur on faster timescales than the qubit dynamics, the cavity degrees of freedom can be eliminated from the dynamics. Starting from  Eq.~(8) of the main paper for $n=3$, we obtain an effective Hamiltonian acting solely on the three-qubit Hilbert space. In a compact form, the effective Hamiltonian can be written as
\begin{equation}
H = H_{\mathrm{q}} + H_{\mathrm{cs}} + H_{J} + H_{\mathrm{qc}} + H_{\mathrm{qd}} + H_{\mathrm{cd}},
\label{eq:Heff_3q}
\end{equation}
where $H_{\mathrm{q}}$ is the free qubit Hamiltonian, $H_{\mathrm{cs}}$ is the cavity-induced (Stark-shift-like) contribution, $H_J$ is the effective qubit-qubit interaction mediated by the cavity network, and $H_{\mathrm{qc}}$, $H_{\mathrm{qd}}$, $H_{\mathrm{cd}}$ are, respectively, the residual qubit-cavity coupling and the direct qubit and cavity driving terms that survive elimination.
The free qubit Hamiltonian and the direct qubit drive are:
\begin{equation}
H_{\mathrm{q}} = \sum_{j=1}^{3} \frac{\Delta_j}{2}\sigma_z^{(j)},
\qquad
H_{\mathrm{qd}} = \sum_{j=1}^{3}\frac{\epsilon_{q_j}}{2} \sigma_x^{(j)}.
\label{eq:Hq_Hqd_3q}
\end{equation}
To carry out the elimination, we collect the qubit-dependent source terms into the vector
\begin{equation}
\mathbf{X} =
\begin{pmatrix}
X_1\\
X_2\\
X_3
\end{pmatrix},
\qquad
X_j=g_j\sigma_-^{(j)}+\epsilon_c,
\label{eq:X_3q}
\end{equation}
and introduce
\begin{equation}
D=\kappa+i\Delta_c.
\label{eq:D_3q}
\end{equation}
The cavity fields reach steady state according to
\begin{equation}
\mathbf{A}\mathbf{a}=-i\mathbf{X},
\label{eq:A_system_3q}
\end{equation}
with
\begin{equation}
\mathbf{A} =
\begin{pmatrix}
D & iJ_{12} & iJ_{31}\\
iJ_{12} & D & iJ_{23}\\
iJ_{31} & iJ_{23} & D
\end{pmatrix},
\qquad
\mathbf{a} =
\begin{pmatrix}
a_1\\
a_2\\
a_3
\end{pmatrix},
\label{eq:A_3q}
\end{equation}
whose determinant is
\begin{equation}
\mathcal{D} = D^3 + D\left(J_{12}^2+J_{23}^2+J_{31}^2\right) - 2iJ_{12}J_{23}J_{31}.
\label{eq:Dcal_3q}
\end{equation}
Inverting Eq.~\eqref{eq:A_system_3q} then gives the eliminated cavity operators as
\begin{equation}
\mathbf{a} = -\frac{i}{\mathcal{D}}\mathbf{B},
\qquad
\mathbf{B} = \operatorname{adj}(\mathbf{A})\mathbf{X},
\label{eq:a_B_3q}
\end{equation}
with
\begin{equation}
\mathbf{B} =
\begin{pmatrix}
B_1\\
B_2\\
B_3
\end{pmatrix}.
\label{eq:B_3q}
\end{equation}
Written out, the three components are
\begin{align}
B_1={}& (D^2+J_{23}^2)X_1 -(iJ_{12}D+J_{31}J_{23})X_2 \nonumber\\
&-(J_{12}J_{23}+iJ_{31}D)X_3,
\label{eq:B1_3q}
\\
B_2={}& -(iJ_{12}D+J_{31}J_{23})X_1 +(D^2+J_{31}^2)X_2 \nonumber\\
&-(J_{12}J_{31}+iJ_{23}D)X_3,
\label{eq:B2_3q}
\\
B_3={}& -(J_{12}J_{23}+iJ_{31}D)X_1 -(J_{12}J_{31}+iJ_{23}D)X_2 \nonumber\\
&+(D^2+J_{12}^2)X_3.
\label{eq:B3_3q}
\end{align}
With these, the cavity-induced contribution to the effective Hamiltonian takes the form
\begin{equation}
H_{\mathrm{cs}} = \frac{\Delta_c}{|\mathcal{D}|^2} \sum_{j=1}^{3}B_j^\dagger B_j,
\label{eq:Hcs_3q}
\end{equation}
and the interaction generated by virtual photon exchange through the cavity network is
\begin{equation}
H_J
=
\frac{1}{|\mathcal{D}|^2}
\sum_{i<j} J_{ij}
\left(
B_i^\dagger B_j+B_j^\dagger B_i
\right),
\label{eq:HJ_general}
\end{equation}
The remaining pieces associated with the eliminated cavity fields are the residual qubit-cavity coupling
\begin{equation}
H_{\mathrm{qc}} = \sum_{j=1}^{3} g_i \left( Y_j^\dagger\sigma_-^{(j)} + Y_j\sigma_+^{(j)} \right),
\qquad
Y_j=-\frac{iB_j}{\mathcal{D}},
\label{eq:Hqc_3q}
\end{equation}
and the cavity drive term
\begin{equation}
H_{\mathrm{cd}} = \epsilon_c \sum_{j=1}^{3} \left( Y_j+Y_j^\dagger \right).
\label{eq:Hcd_3q}
\end{equation}
Equations~\eqref{eq:Heff_3q}-\eqref{eq:Hcd_3q} together constitute the effective three-qubit Hamiltonian obtained after tracing out the cavity modes. It combines local qubit terms with cavity-mediated many-body interactions; notably, since each $B_i$ depends on all three lowering operators, the induced interaction is genuinely nonlocal across the qubit manifold, even though every qubit couples only to its own cavity mode at the microscopic level.

The spin-spin coupling in (\ref{eq:Hcs_3q}) - (\ref{eq:Hqc_3q}) together can be mapped as in the Kuramoto model, in which the coupling constants are of the form $\cos(\phi_i-\phi_j)$, with $\phi_{i,j}$ being azimuthal angles in the Bloch sphere of the $i$th and $j$th spins, respectively. As in the two-spin case, the local spin driving and local spin dissipation and dephasing make it deviate from the classical Kuramoto model, in addition to the quantum framework (ref.~[8] in main text).

The Hamiltonian in Eq.~\eqref{eq:Heff_3q} enters the Lindblad master equation used to evolve the three-qubit density matrix,
\begin{equation}
\frac{d\rho}{dt}
=
-i\left[H_{\mathrm{eff}},\rho\right]
+
\sum_{j=1}^{3}
\gamma_j\mathcal{L}
\left[\sigma_-^{(j)}\right]\rho
+
\sum_{j=1}^{3}
\gamma_{\phi j}\mathcal{L}
\left[\sigma_z^{(j)}\right]\rho ,
\label{eq:master_3q}
\end{equation}
where the dissipator is defined as usual,
\begin{equation}
\mathcal{L}[M]\rho
=
M\rho M^\dagger
-\frac{1}{2}
\left\{
M^\dagger M,\rho
\right\}.
\label{eq:dissipator_3q}
\end{equation}

\section{Measure of tripartite synchronization}

We start with a density matrix $\rho_{123}$ characterized by the phase variables $\phi_j\in[0,2\pi)$, $j=1,2,3$. 
For each pair $(i,j)$ of these spins, we introduce the relative phase $\phi_-^{(ij)}= (\phi_i-\phi_j)/2$,
together with the corresponding collective phase
$\phi_+^{(ij)}=(\phi_i+\phi_j)/2,$ so that $\phi_i=\phi_+^{(ij)}+\phi_-^{(ij)},
\phi_j=\phi_+^{(ij)}-\phi_-^{(ij)}$.
For the corresponding reduced density matrix $\rho_{ij}$, the joint Husimi function can be written as 
\begin{equation}
Q_{ij}(\theta,\phi_+^{(ij)},\phi_-^{(ij)})=\frac{1}{4\pi^2}
\langle\theta,\phi_+^{(ij)},\phi_-^{(ij)}|
\rho_{ij}
|
\theta,\phi_+^{(ij)},\phi_-^{(ij)}
\rangle.
\label{eq:trip_Qij_full}
\end{equation}
where $|\theta,\phi_+^{(ij)},\phi_-^{(ij)}\rangle\equiv|\theta,\phi_i\rangle\otimes|\theta,\phi_j\rangle.$ is the two-spin coherent state, with $|\theta,\phi_j\rangle$ being the single-spin coherent state [see main text].
Projecting out the collective phase, we get [see Eq. (10) in the main text]
\begin{equation}
Q_{ij}(\theta,\phi_-^{(ij)})=2\int_{|\phi_-^{(ij)}|}^{2\pi-|\phi_-^{(ij)}|}Q_{ij}(\theta,\phi_+^{(ij)},\phi_-^{(ij)})
\,d\phi_+^{(ij)}.
\label{eq:trip_Qij}
\end{equation}
The corresponding relative-phase distribution then becomes [see Eq. (9) in the main text]
\begin{equation}
P_{ij}(\phi_-^{(ij)})=\int_0^\pi Q_{ij}(\theta,\phi_-^{(ij)})
\sin\theta\,d\theta,
\label{eq:trip_Pij}
\end{equation}
and the pairwise synchronization measure is
\begin{equation}
R_{ij}=\left|\int_{-\pi}^{\pi}P_{ij}(\phi_-^{(ij)})e^{i\phi_-^{(ij)}}\,d\phi_-^{(ij)}\right|.
\label{eq:trip_Rij}
\end{equation}

We next introduce two independent relative phases $\phi_-^{(12)}=(\phi_1-\phi_2)/2$ and $\phi_-^{(23)}
=(\phi_2-\phi_3)/2$, and define $\Phi_\pm=(\phi_-^{(12)}\pm\phi_-^{(23)})/2.$ Consequently, $\Phi_- =(\phi_1-2\phi_2+\phi_3)/4$ represents a collective three-body relative-phase coordinate. The inverse transformation therefore reads as $\phi_1=\phi_2+c_+$, $\phi_3=\phi_2+c_-$, where $c_\pm= 2(\Phi_+\pm \Phi_-)$.
Hence, the projected tripartite Husimi distribution in $\Phi_-$ space is obtained by integrating over the other phase coordinates with the constraint 
$\phi_1,\phi_2,\phi_3\in[0,2\pi)$:
\begin{equation}
\begin{aligned}
Q_{123}(\theta,\Phi_-)
&=8\int_{\mathcal{D}_+}d\Phi_+
\int_{\mathcal{D}_2} d\phi_2\,
Q_{123}\left(\theta,\phi_2+c_+,\phi_2,
\phi_2+c_-\right),
\end{aligned}
\label{eq:Q123_projected}
\end{equation}
where $Q_{123}$ is joint three-spin Husimi-$Q$ function given by
\begin{equation}
Q_{123}(\theta,\phi_1,\phi_2,\phi_3)=\frac{1}{(2\pi)^3}
\langle \theta,\Phi_+,\Phi_-,\phi_2|\rho_{123}|\theta,\Phi_+,\Phi_-,\phi_2\rangle
\label{eq:Q123_full}
\end{equation}
and 
\begin{equation}
|\theta,\Phi_+,\Phi_-,\phi_2\rangle\equiv|\theta,\phi_1\rangle
\otimes|\theta,\phi_2\rangle\otimes|\theta,\phi_3\rangle.
\label{eq:trip_coherent}
\end{equation}
is the three-spin coherent state. The factor of 8 appears in (\ref{eq:Q123_projected}) as the Jacobian for the coordinate transformation $(\phi_1,\phi_2,\phi_3)
\rightarrow
(\Phi_+,\Phi_-,\phi_2)$.

Note that the two-variable integration in (\ref{eq:Q123_projected}) depends upon relevant domains $\mathcal{D}_+$ and $\mathcal{D}_2$ which can be identified using the constraint  $\phi_1,\phi_2,\phi_3\in[0,2\pi)$ as follows:
\begin{equation}
\mathcal{D}_2:\quad
\phi_2\in\big[\,L(\Phi_+,\Phi_-),\,U(\Phi_+,\Phi_-)\,\big],
\label{eq:phi2_domain}
\end{equation}
with
\begin{equation}
\begin{aligned}
L &= \max(0,-c_+,-c_-),\\
U & 
   = 2\pi+\min(0,-c_+,-c_-).
\end{aligned}
\label{eq:LU_def}
\end{equation}
The length of this interval is
\begin{equation}
\begin{aligned}
\ell(\Phi_+,\Phi_-)&=U-L\\
&=2\pi-\Big[\max(0,-c_+,-c_-)-\min(0,-c_+,-c_-)\Big],
\label{eq:phi2_length}
\end{aligned}
\end{equation}
which becomes nonnegative when 
\begin{equation}
|\Phi_+|<\min\!\left(\frac{\pi}{2},\,\pi-|\Phi_-|\right).
\label{eq:Phi_plus_condition}
\end{equation}
This defines the domain $\mathcal{D}_+$ for the integration with respect to $\Phi_+$:
\begin{equation}
\mathcal{D}_+:\quad
\Phi_+\in
\begin{cases}
\left(-\dfrac{\pi}{2},\dfrac{\pi}{2}\right), & |\Phi_-|\le \dfrac{\pi}{2},\\[2mm]
\big(|\Phi_-|-\pi,\ \pi-|\Phi_-|\big), & \dfrac{\pi}{2}<|\Phi_-|<\pi.
\end{cases}
\label{eq:Phi_plus_domain}
\end{equation}

The distribution function for the relative collective phase $\Phi_-$  is then obtained by integrating over the polar angle $\theta$:
\begin{equation}
P_{123}(\Phi_-)
=
\int_0^\pi
Q_{123}(\theta,\Phi_-)
\sin\theta\,d\theta,
\label{eq:P123}
\end{equation}
with
\begin{equation}
\int_{-\pi}^{\pi}
P_{123}(\Phi_-)\,d\Phi_-
=
1.
\end{equation}
The tripartite synchronization measure is consequently defined from the first circular moment of this collective relative-phase distribution:
\begin{equation}
R_{123}
=\left|\left\langle e^{i\Phi_-}\right\rangle\right|=
\left|
\int_{-\pi}^{\pi}
P_{123}(\Phi_-)
e^{i\Phi_-}\,d\Phi_-
\right|.
\label{eq:R123}
\end{equation}
which describes the localization of the three-body relative phase
$\Phi_-$, while $R_{ij}$ probes pairwise relative-phase localization and $R_i$ characterizes the localization of an individual phase.
The above construction provides a direct quantum-state-based route from the three-spin density matrix to the single-, pair-, and tripartite phase-localization measures. In particular, the quantities $R_i$, $R_{ij}$, and $R_{123}$ are all obtained from the first circular moments of phase distributions generated from the corresponding Husimi-$Q$ functions.